\documentclass[11pt, twoside]{article}
\usepackage{amssymb}

\newcommand{\bea}{\begin{eqnarray*}}
\newcommand{\eea}{\end{eqnarray*}}
\newcommand{\alfa}{\alpha}
\newcommand{\teta}{\theta}
\newcommand{\fie}{\varphi}
\newcommand{\Fie}{\Phi}

\newtheorem{teorema}{Theorem}
\newtheorem{vermoeden}{Conjecture}
\newtheorem{lemma}{Lemma}

\begin{document}

\title{Conjectured $DXZ$ decompositions \\ of a unitary matrix}

\author{Alexis De Vos, Martin Idel, and Stijn De Baerdemacker}
\maketitle

\begin{abstract}
For any unitary matrix there exists a ZXZ decomposition,
according to a theorem by Idel and Wolf.
For any even-dimensional unitary matrix there exists a block-ZXZ decomposition,
according to a theorem by F\"uhr and Rzeszotnik.
We conjecture that these two decompositions are merely special cases of 
a set of decompositions, one for every divisor of the matrix dimension.
For lack of a proof, we provide an iterative Sinkhorn algorithm
to find an approximate numerical decomposition.
\end{abstract}



\section{Introduction}

Recently, two decompositions of an arbitrary $n \times n$ unitary matrix~$U$ 
into a matrix product $DXZ$ of three unitary matrices have been proposed:
\begin{itemize}
\item For arbitrary $n$, Idel and Wolf \cite{idel} 
      present a decomposition 
      where $D$~and~$Z$ are diagonal matrices, 
      whereas $X$ is a matrix with all linesums equal to unity.  
\item For arbitrary even $n$, F\"uhr and Rzeszotnik \cite{fuhr} 
      present a decomposition
      where $D$ and $Z$ are block-diagonal matrices, 
      whereas $X$ is a matrix with all block-linesums equal to 
      the $n/2 \times n/2$ unit matrix.  
\end{itemize}
These matrix decompositions have been applied in
quantum optics    \cite{idel}, 
quantum computing \cite{freiberg} 
                  \cite{devosdebaerdemacker} \cite{vanrentergem}, and 
quantum memory    \cite{simnacher}.

The two matrix decompositions have been proved in a very different way.
Whereas 
the proof of the Idel--Wolf         decomposition 
(based on symplectic topology) is not constructive, 
the proof of the F\"uhr--Rzeszotnik decomposition 
(based on linear algebra)      is     constructive.
In the present paper, we conjecture that nevertheless
the two decompositions belong to a same set of similar decompositions.
We conjecture that there exist as many such decompositions
as there are divisors of the number~$n$.
We present no proof, as 
neither the Idel--Wolf         proof 
nor     the F\"uhr--Rzeszotnik proof can be easily extrapolated.  

\section{Conjecture}

We introduce the following three positive integers:
\begin{itemize}
\item $n$, an arbitrary integer greater than~1,
\item $m$, a divisor of~$n$, 
           distinct\footnote{The restriction $m \neq n$ is merely
                             introduced for convenience.
                             The reader may easily investigate 
                             the case $m=n$. E.g., if $m=n$,
                             then Conjecture~1 is trivially true:
                             suffice it to choose $D$ equal to $U$
                             and both $X$ and     $Z$ equal to 
                             the $n \times n$ unit matrix.} 
           from~$n$, and
\item $q$, equal to~$n-m$.
\end{itemize}
We write $n=rm$ and $q=(r-1)m$.
Hence, both $m$ and $r$ are divisors of~$n$. 
They satisfy $1 \le m < n$ and $1 < r \le n$.
For convenience, 
$n \times n$ matrices will be called `great        matrices',
$m \times m$ matrices will be called `small        matrices', and
$q \times q$ matrices will be called `intermediate matrices'.

\begin{vermoeden}
Every great unitary matrix~$U$ can be decomposed into
three great unitary matrices:
\[
U = DXZ \ ,
\]
where
\begin{itemize} 
\item $D$ consists of $r$ small matrices on its diagonal:
      \[ 
      D = \mbox{diag}\, (D_{11}, D_{22}, D_{33}, ..., D_{rr}) \ ,
      \]
\item $Z$ consists of $r$ small matrices on its diagonal,
      the upper-left small matrix being equal to 
      the $m \times m$ unit matrix~$I$:
      \[ 
      Z = \mbox{diag}\, (I, Z_{22}, Z_{33}, ..., Z_{rr}) \ ,
      \] and
\item $X$ consists of $r^2$ small matrices $X_{jk}$, such that 
      all row    sums $\sum_{k=1}^rX_{jk}$ and
      all column sums $\sum_{j=1}^rX_{jk}$ are equal to
      the small matrix~$I$.
\end{itemize}      
\end{vermoeden}
Because $D$ is unitary, 
automatically all its blocks $D_{jj}$ are unitary; 
because $Z$ is unitary, 
automatically all its blocks $Z_{jj}$ are unitary.
In contrast, the blocks $X_{jk}$ are not necessarilly unitary.

We define the $n \times n$ transformation matrix
\[
T = F_r \otimes I \ ,
\]
where the matrix $F_r$ is 
the $r \times r$ discrete Fourier transform.
We can easily demonstrate that the product 
$T^{-1}XT$ is of the form
\[
\left( \begin{array}{cc} I & \\ & G \end{array} \right) \ . 
\]
We thus have the following property:
\[
X = T \left( \begin{array}{cc} I & \\ & G \end{array} \right) T^{-1} \ .
\] 
Because both $X$ and $T$ are unitary, 
automatically $G$ is a unitary intermediate matrix.

We summarize that the decomposition of $U$ 
corresponds to finding 
the      appropriate $2r-1$ small        unitary matrices and 
a single appropriate        intermediate unitary matrix.
This corresponds to find the appropriate
\[
(2r-1)m^2 + q^2 = (2r-1)m^2 + [(r-1)m]^2 = r^2m^2
\]
real parameters, a number which exactly matches $n^2$,
i.e.\ the number of degrees of freedom of the given matrix~$U$. 

\section{Three special cases}

If $m=1$ (and thus $q=n-1$), then all small matrices are, 
in fact, just complex numbers.
Both $D$ and $Z$ are diagonal unitary matrices
and $X$ is a unit linesum     unitary matrix.
The transformation matrix~$T$ equals the great Fourier matrix $F_n$.
In this particular case, the above conjecture
has been proposed by De Vos and De Baerdemacker \cite{devos}
and subsequently proved by Idel and Wolf \cite{idel}.
The proof is by symplectic topology.
Unfortunately, the proof is not constructive
and therefore only provides the guarantee
that the numbers $D_{11}, D_{22}, ..., D_{nn}$ and
$Z_{22}, Z_{33}, ..., Z_{nn}$ exist, without providing their values.
De Vos and De Baerdemacker \cite{devos} give a 
Sinkhorn algorithm that yields 
numerical approximations of these numbers.
Finally, we note that examples of the case $m=1$ demonstrate that the $DXZ$
decomposition is not always unique.

If $n$ is even and $m$ equals $n/2$ (and thus $q=n/2$), 
then intermediate matrices are,
in fact, small matrices.
The transformation matrix~$T$ equals $H \otimes I$,
where $H=F_2$ is the $2 \times 2$ Hadamard matrix.
In this particular case, the above conjecture 
has been proved by F\"uhr and Rzeszotnik \cite{fuhr}.
The proof is constructive and thus gives
explicit values for the small matrices $D_{11}$, $D_{22}$, $Z_{22}$, and $G$.
Also in this special case decomposition 
is not unique \cite{devosdebaerdemacker} \cite{rzeszotnik}.
 
Finally, if both $m=1$ and $m=n/2$, i.e. if $n=r=2$ and $m=q=1$,
then the decomposition is well-known.
An arbitrary matrix from U(2) looks like
\begin{equation}
U = \left( \begin{array}{rl} \cos(\fie)e^{i(\teta+\psi)} & 
                             \sin(\fie)e^{i(\teta+\chi)} \\ 
                            -\sin(\fie)e^{i(\teta-\chi)} & 
                             \cos(\fie)e^{i(\teta-\psi)} \end{array} \right) \ .
\label{u2}
\end{equation}
One possible decomposition is
\begin{equation}
U = \left( \begin{array}{cc} e^{i(\teta+\fie+\psi)} & \\ & ie^{i(\teta+\fie-\chi)} \end{array} \right) 
    \ \frac{1}{2}\ 
    \left( \begin{array}{cc} 1 + e^{-2i\fie} & 1 - e^{-2i\fie} \\
                             1 - e^{-2i\fie} & 1 + e^{-2i\fie} \end{array} \right)
    \left( \begin{array}{cc} 1 & \\ & -ie^{i(-\psi+\chi)}\end{array} \right) \ .
\label{xu2}
\end{equation}

\section{Group hierarchy}

The matrices $D$ form a group isomorphic to U($m$)$^r$,
of dimension $rm^2 = nm$.
The matrices $Z$ form a group isomorphic to U($m$)$^{r-1}$,
of dimension $(r-1)m^2 = (n-m)m$. Finally, 
the matrices $X$ form a group isomorphic to  U($q$),
of dimension $q^2 = (n-m)^2$.
We denote these three matrix groups by 
DU($n,m$), ZU($n,m$), and XU($n,m$), respectively.    
In particular, the groups XU($n, 1$) and ZU($n, 1$) 
are the groups XU($n$) and ZU($n$), 
extensively studied in the past \cite{vanrentergem} \cite{debaerdemacker}.

According to the conjecture, for any~$m$, 
the closure of the groups DU($n,m$), ZU($n,m$), and XU($n,m$)
is the unitary group of $U$~matrices.
Of course, because ZU($n,m$) is a subgroup of DU($n,m$),
the closure of        DU($n,m$) and XU($n,m$) is also U($n$). In fact, 
the closure of merely ZU($n,m$) and XU($n,m$)
already equals U($n$).
Indeed, any DU($n,m$) matrix can be decomposed into 
two ZU($n,m$) matrices and
two XU($n,m$) matrices:
\bea
&&
\left( \begin{array}{ccccc} D_{11} &        &        &        &        \\
                                   & D_{22} &        &        &        \\
                                   &        & D_{33} &        &        \\
                                   &        &        & \ddots &        \\
                                   &        &        &        & D_{rr} \end{array} \right) =
\\ && \hspace{-20mm}
\left( \begin{array}{ccccc}        &   I    &        &        &        \\
                              I    &        &        &        &        \\
                                   &        &    I   &        &        \\
                                   &        &        & \ddots &        \\
                                   &        &        &        &    I   \end{array} \right)\left( \begin{array}{ccccc}   I    &        &        &        &        \\
                                   & D_{11} &        &        &        \\
                                   &        &    I   &        &        \\
                                   &        &        & \ddots &        \\
                                   &        &        &        &    I   \end{array} \right)\left( \begin{array}{ccccc}        &   I    &        &        &        \\
                              I    &        &        &        &        \\
                                   &        &   I    &        &        \\
                                   &        &        & \ddots &        \\
                                   &        &        &        &    I   \end{array} \right)\left( \begin{array}{ccccc}   I    &        &        &        &        \\
                                   & D_{22} &        &        &        \\
                                   &        & D_{33} &        &        \\
                                   &        &        & \ddots &        \\
                                   &        &        &        & D_{rr} \end{array} \right) \ .
\eea  

If $m>1$, then we have the following group hierarchy:
\bea
\mbox{U}(n) \supset & \hspace*{-3mm} \mbox{DU}(n,m) \supset \mbox{DU}(n,1) = \mbox{DU}(n) \hspace*{-2mm} & \\                   
                    & \hspace*{-3mm} \mbox{XU}(n,m) \subset \mbox{XU}(n,1) = \mbox{XU}(n) \hspace*{-2mm} & \subset \mbox{U}(n) \ .
\eea
In fact, we have the following isomorphisms:
\bea
\mbox{DU}(n,m) & \cong & \mbox{U}(m)^{n/m}   \\
\mbox{ZU}(n,m) & \cong & \mbox{U}(m)^{n/m-1} \\
\mbox{XU}(n,m) & \cong & \mbox{U}(n-m)       \ .
\eea
If $n$ is a power of a prime, say $p^w$,
then $m$ necessarilly is also a prime power, say~$p^u$ (with $0 \le u < w$).
The XU($n,m$) groups with all possible values of~$m$ (i.e.\ $1, p, p^2, ..., p^{w-1}$)
form an elegant subgroup chain according to
\[
\mbox{XU}(p^w) =
\mbox{XU}(p^w, 1) \supset \mbox{XU}(p^w, p) \supset \mbox{XU}(p^w, p^2) 
                  \supset ...               \supset \mbox{XU}(p^w, p^{w-1}) \ ,
\]  
with successive dimensions
\[
(p^w-1)^2 > (p^w-p)^2 > (p^w-p^2)^2 > ... > (p^w-p^{w-1})^2  \ .
\]

\section{Conjugate conjecture}

If Conjecture 1 is true, 
then automatically a second conjecture is also true:
\begin{vermoeden}
Every great unitary matrix~$U$ can be decomposed into
three great unitary matrices:
\[
U = C \left( \begin{array}{cc} I & \\ & A \end{array} \right) Y \ ,
\]
where
\begin{itemize} 
\item $C$ is a circulant $n \times n$ matrix,
      i.e.\ a unitary matrix consisting of $m \times m$ small blocks,
      such that two $C_{jk}$ are identical if their two $j-k$ are equal,
\item $A$ is a  $q \times q$ unitary   matrix, and
\item $Y$ is an $n \times n$ circulant matrix, 
      the upper row sum\footnote{As the matrix is circulant, 
                                 all row sums and column sums are equal.
                                 Hence $Y$ is a member of XU($n, m$).}
      being equal to the $m \times m$ unit matrix~$I$.
\end{itemize}      
\end{vermoeden}
Indeed, if we apply Conjecture~1, not to the given matrix~$U$,
but instead to its conjugate
\[
u = T^{-1} U T \ ,
\]
then we obtain the decomposition
\[
u = dxz \ .
\]
This leads to
\[
U = T\, d\, T^{-1}T\, x\, T^{-1}T\, z T^{-1} \ .
\]
One can easily verify that
\begin{itemize} 
\item $T\, d\, T^{-1}$ is a circulant great matrix,
\item $T\, x\, T^{-1}$ is of the form $\left( \begin{array}{cc} I & \\ & A \end{array} \right)$, and 
\item $T\, z\, T^{-1}$ is a circulant XU($n, m$) matrix.
\end{itemize}
Such conjugate decomposition was already noticed before,
in both the $m=1$ case and the $m=n/2$ case 
\cite{idel} \cite{vanrentergem} \cite{debaerdemacker}. 

\section{Unitary and biunitary vectors}

For $m=1$, the DXZ decomposition involves 
unit-modulus numbers~$d_{jj}$ and~$z_{jj}$:
\[
U = \left( \begin{array}{rrrr} d_{11} &        &        &       \\
                                      & d_{22} &        &       \\
                                      &        & \ddots &       \\
                                      &        &        & d_{nn} 
           \end{array} \right) X
    \left( \begin{array}{rrrr}      1 &        &        &       \\

                                      & z_{22} &        &       \\

                                      &        & \ddots &       \\
                                      &        &        & z_{nn} 

           \end{array} \right) \ .
\]
If we multiply both sides of the equation
by the $n \times 1$ matrix (i.e.\ column vector)
$v = (1, z_{22}^{-1}, z_{33}^{-1}, ..., z_{nn}^{-1})^T$,
then we obtain
\[
Uv  = \left( \begin{array}{rrrr} d_{11} &        &        &       \\
                                        & d_{22} &        &       \\
                                        &        & \ddots &       \\
                                        &        &        & d_{nn} 
             \end{array} \right) X
      \left( \begin{array}{r} 1 \\ 1 \\ \vdots \\ 1 \end{array} \right) \ .
\] 
Taking into account that all row sums of $X$ equal unity,
we find
\[
Uv = w \ ,
\]
where $w = (d_{11}, d_{22}, d_{33}, ..., d_{nn})^T$.
Both $v$ and $w$ are vectors with all entries having unit modulus.
Therefore, they are called unimodular vectors.
The unimodular vector $v$ is
called biunimodular for the matrix~$U$, 
as $Uv$ is unimodular as well \cite{idel} \cite{fuhr}.
We say that the Idel--Wolf DXZ decomposition implies 
the fact that any unitary matrix 
has at least one biunimodular vector.
Moreover, it possesses a biunimodular vector with leading entry~1.
As an example, decomposition (\ref{xu2}) of the matrix (\ref{u2})
corresponds with the following biunimodular vector:
\[
U\ \left( \begin{array}{c} 1 \\ i\, e^{i(\psi - \chi)}
          \end{array} \right) =
   \left( \begin{array}{r} e^{i(\fie + \teta + \psi)} \\ 
                       i\, e^{i(\fie + \teta - \chi)}
          \end{array} \right) \ .
\]

If Conjecture~1 is true for $m>1$,
then we can draw a similar conclusion $UV=W$,
however with $V$ and $W$ matrices of size $n \times m$.
These matrices consist of $r$ blocks, each a unitary $m \times m$ matrix.
Because such blocks have no modulus, 
we cannot call $V$ and $W$ unimodular vectors.
We will instead call them unitary vectors
and $V$ a biunitary vector.
These unitary vectors reside in an $nm$-dimensional vector space 
$\mathbb{C}^{n}\otimes\mathbb{C}^{m}$, isomorphic to
$\mathbb{R}^{2nm}$. 
A~basis for this space consists e.g.\ of the $nm$~following basis vectors:
$a_i \otimes b_j^T$, where 
the $a_i$ are the $n$~standard basis vectors of $\mathbb{C}^n$ and
the $b_j$ are the $m$~standard basis vectors of $\mathbb{C}^m$.
We note that a unitary vector~$V$ has the property $V^{\dagger}V=rI$,
with $I$ once again the $m \times m$ unit matrix. 

If Conjecture~1 is true, then also the following conjecture is true:
\begin{vermoeden}
Every great unitary matrix~$U$ has at least one biunitary vector~$V$:
\[
UV = W ,
\]
where
\begin{itemize} 
\item both $V$ and $W$ consist of $n/m$ unitary $m \times m$ entries and
\item $V$ has leading entry equal to the small unit matrix~$I$.
\end{itemize}      
\end{vermoeden}
Suffice it to repeat the above reasoning with $m=1$ for $m>1$,  
the vector $E = (I, I, I, ..., I)^T$ taking over the role of
the vector $e = (1, 1, 1, ..., 1)^T$ above.

Important is the fact that not only
Conjecture~3 is         a consequence of Conjecture~1, but
Conjecture~1 is equally a consequence of Conjecture~3.
Indeed, if $UV = W$, with both $V$ and $W$ being unitary vectors
and $V$ having the unit matrix~$I$ as leading entry, 
then the matrix
\[
A = \mbox{diag}\, (W_1^{-1}, W_2^{-1}, ..., W_r^{-1}) \ U \ 
    \mbox{diag}\, (       I, V_2^{-1}, ..., V_r^{-1})
\]
belongs to XU($n$, $m$).
Proof of this fact consists of two parts:
\begin{itemize}
\item Taking into account that $UV = W$, we find that $AE$ equals $E$,
      such that $A$ has unit row sums.
\item Because of $E=AE$, we have 
      $E = \overline{E} = \overline{AE} = \overline{A}E$.
      Taking into account that $A$ is unitary, 
      we have $A^T\,\overline{A}$ equal to the $n \times n$ unit matrix.
      Hence $E=A^T\,\overline{A}E =A^TE$.
      Because $A^TE$ thus turns out to equal~$E$, 
      we conclude that $A$ has unit column sums.
\end{itemize}
As~$A$ belongs to XU($n$, $m$) and $U$ has decomposition
\[
 \mbox{diag}\, (W_1, W_2, ..., W_r) \ A \ 
 \mbox{diag}\, (  I, V_2, ..., V_r) \ ,
\]
Conjecture~1 is fulfilled.    

We finally note that Conjecture~2 leads to the same Conjecture~3,
according to a similar proof, where however
the vector $  (I, 0, 0, ..., 0)^T$ takes over the role of
the vector $E=(I, I, I, ..., I)^T$ above.
We conclude:
\begin{teorema} 
The Conjectures~1, 2, and 3 are equivalent:
if one is proved, then all three are proved.
\end{teorema}  

\section{Group topology}

In order to prove the three conjectures,
it suffices to prove Conjecture~3.
For that purpose, we first give a lemma:
\begin{lemma}
If an $n \times n$ unitary matrix $U$ possesses
a biunitary $n \times m$ vector,
then it possesses   a biunitary $n \times m$ vector
with leading entry equal to the $m \times m$ unit matrix~$I$.
\end{lemma}
Indeed, let us suppose that
\[
U \left( \begin{array}{c} V_1 \\ V_2 \\ V_3 \\ \vdots \\ V_r 
           \end{array} \right) =
  \left( \begin{array}{c} W_1 \\ W_2 \\ W_3 \\ \vdots \\ W_r 
           \end{array} \right) \ ,
\]
with all $V_j$ and $W_j$ are unitary blocks.
We multiply to the right with the small matrix $V_1^{-1}$
and thus obtain
\[
U \left( \begin{array}{c} I \\ V_2V_1^{-1} \\ V_3V_1^{-1} \\ 
                                    \vdots \\ V_rV_1^{-1} 
           \end{array} \right) =
  \left( \begin{array}{c} W_1V_1^{-1} \\ W_2V_1^{-1} \\ W_3V_1^{-1} \\ 
                                              \vdots \\ W_rV_1^{-1} 
           \end{array} \right) \ ,
\]
a result which proves the lemma.

We consider the vector space ${\bf M}$ of vectors $(M_1, M_2, ..., M_r)^T$,
where each $M_j$ is a complex $m \times m$ matrix.
Let ${\bf S}$ be the the following submanifold of~${\bf M}$:
\[
{\bf S} = \{ (V_1, V_2, ..., V_r)^T\ |\  V_j \in {\mbox U}(m) \} \ . 
\]
The Lie group U($m$)$^r$ behaves as if it were
the following topological product of 
odd-dimensional spheres~\cite{pontrjagin} \cite{samelson}:
\[
\left( S^1 \times S^3 \times S^5 \times ... \times S^{2m-1}\right)^r \ ,
\]
where $S^k$ denotes the $k$-sphere.
In fact, the Poincar\'e polynomial of the manifold $\bf S$ is
\[
P(x) = [\ (1+x)(1+x^3)(1+x^5)...(1+x^{2m-1})\ ]^r \ .
\]
Therefore, the sum of its Betti numbers is
\[
P(1) = (2^m)^r = 2^n \ ,
\] 
where $n=mr$.

It is clear that, if 
\begin{equation}
{\bf S} \cap\, U{\bf S} \neq \emptyset \ ,
\label{empty}
\end{equation}
then there exists at least one unitary vector in ${\bf S}$
which is a biunitary vector an which, because of Lemma~1, 
has a unit leading entry.

One promising approach is to reduce the problem to the Arnold conjecture \cite{arnold}, 
as has been done in the $m=1$ case \cite{idel}.
If ${\bf S}$ was a Lagrangian submanifold for a symplectic form on $\mathbb{C}^n$ 
such that $U$ was still a Hamiltonian symplectomorphism, 
then eqn (\ref{empty}) would be true, provided the Arnold conjecture 
is true for this particular manifold.

Direct computation suggests that ${\bf S}$ is no Lagrangian submanifold of $\mathbb{C}^n$ 
with the standard symplectic form. 
There are two possible roads to still prove a relation to the Arnold conjecture:
\begin{itemize}
\item show that ${\bf S}$ is a submanifold for some other symplectic structure 
      on $\mathbb{C}^n$ and $U$ is a Hamiltonian symplectomorphism 
      for that particular structure, too
\item find a Lagrangian embedding of ${\bf S}$ into some other manifold 
      such that the mapping of $U$ results in a Hamiltonian symplectomorphism 
      for this other manifold.
\end{itemize}

Let us start with the first idea: we note that ${\bf S}$ is a Cartesian product 
of odd-dimensional spheres and that the Cartesian product of Lagrangian manifolds 
is a Lagrangian manifold, 
it might be possible to consider each sphere separately.
However this is not true, as no sphere $S^n$ with $n>1$ can be embedded into $\mathbb{C}^n$ 
as a Lagrangian manifold according to \cite{gromov}, as no simply-connected manifold 
can be embedded into $\mathbb{C}^n$.
Since ${\bf S}$ is not simply connected as it contains a product factor of $S^1$, 
this does not yet rule out the possibility of finding a symplectic structure 
such that it is a Lagrangian submanifold, but there is no argument we know of.

This leaves us with attempting the second idea: Indeed, using \cite{audin}, 
who attributes this idea to \cite{polterovich}, 
we can find a Lagrangian embedding of every odd-dimensional space via:
\bea
S^{2n+1} & \to & {\bf P}^n(\mathbb{C})\times {\bf P}^n(\mathbb{C}) \\ 
z & \mapsto & (\, [z],[\overline{z}]\, )
\eea
where ${\bf P}^n(\mathbb{C})$ denotes the complex projective space 
and $[z]$ being the canonical projection.
Since $S^1$ is a Lagrangian submanifold of $\mathbb{C}$, 
and products of Lagrangian embeddings are Lagrangian embeddings in the product manifold, 
we can embed ${\bf S}$ as a Lagrangian submanifold.
To be of help, we also would need $U$ to be mapped to a symplectomorphism. 
To do that, we note that, if we decompose 
$z\in {\bf S}$ as $(z_1^1,z_3^1,z_5^1,\ldots,z_{2m+1}^r)$, 
then $U$ acts on any factor $z_i^r$ as $U(0,\ldots,0,z_i^r,0,\ldots,0)$, 
which explains how it must act on $([z],[\overline{z}])$. 
But this implies that $U$ will mix factors of ${\bf P}^n(\mathbb{C})$ 
in our product manifold, 
which in turn results in $U$ not being a symplectomorphism after direct computation.
This does not rule out the second idea either, but shows where the difficulties lie.
It is still unclear whether the applicability of symplectic topology 
to the original problem of a Sinkhorn-like decomposition was a mere coincidence 
or whether there is a deeper link to unitary decompositions 
so it seems worthwile to consider this problem.

We summarize:
if the Arnold conjecture is applicable, 
then the above Conjectures~1, 2, and~3 are true.

\section{Numerical approximation}

We note that the above three conjectures are not constructive.
Only in the case $m=n/2$, do we have explicit expressions
for the matrices $D$, $X$, $Z$, $C$, $A$, and $Y$ and 
for the vectors $V$ and $W$.
For other cases, we can only find numerical approximations.
Therefore, in the present section,
we give a numerical procedure to find,
given the matrix~$U$, 
an approximation of the matrices $D$, $X$, and~$Z$.
It is similar to the Sinkhorn-like method presented earlier for the
$m=1$ case \cite{devos}.

The successive approximations $X_t$ of $X$ are given by
\[
X_0 = U 
\]
and
\[
X_t = L_tX_{t-1}R_t \ .
\]
The diagonal of the left great matrix $L_t$ 
consists of $r$ small matrices $(L_t)_{jj}$,
equal to $\Fie_j^{-1}$, i.e.\  the inverse of the unitary factor 
in the polar decomposition $\Fie_jP_j$ 
of the row sum $r_j=\sum_{k=1}^r(X_{t-1})_{jk}$. 
The right great matrix $R_t$ consists of $r$ small matrices $(R_t)_{kk}$,
equal to $\Upsilon_k^{-1}\Upsilon_1$, 
with $\Upsilon_k^{-1}$ equal to the inverse of the unitary factor 
in the polar decomposition $Q_k\Upsilon_k$ 
of the column sum $c_k=\sum_{j=1}^r(L_tX_{t-1})_{jk}$.
The extra factor $\Upsilon_1$ in the expression of $(R_t)_{kk}$ guarantees
that $(R_t)_{11}$ equals~$I$.
After a sufficient number (say, $\tau$) of iterations,
the product $L = L_{\tau}.L_{\tau-1}...L_1$ and
the product $R = R_1.R_2...R_{\tau}$ yield the desired great matrix~$X$:
\[
X \approx X_{\tau} = LUR \ .
\] 
The fact that all $(R_t)_{11}=I$ guarantees that $R_{11}=I$ and thus that
$R$ belongs to ZU($n,m$) instead of merely to DU($n,m$).
We have
\bea
D & \approx & L^{-1} \\
Z & \approx & R^{-1} \ .
\eea

Exceptionally, a particular row sum $r_j$ might be singular. 
Then its polar decomposition is not unique,
such that the corresponding matrix $\Fie_j$ is not determined.
In that case, we choose $(L_t)_{jj}$ equal to the unit matrix~$I$.
Analogously we choose $(R_t)_{kk} = I$ whenever a particular column sum $c_k$ is singular.

The progress of the iteration process can be monitored 
by the following property of a great matrix~$M$:
\[
\Psi(M) = n^2 - |\mbox{Btr}(M)|^2 
\]
where we call $\mbox{Btr}(M)$ the `block trace' of~$M$:  
\[
\mbox{Btr}(M) = \sum_{j=1}^r \sum_{k=1}^r \,\mbox{Tr} (M_{jk}) \ .
\]
Indeed, the quantity $\Psi(M)$ is zero iff $M \in$ XU($n, m$).
During the iteration process, $\Psi(X_t)$ becomes smaller and smaller,
approaching zero in the limit.
See~Appendix for details.
We note that, if $m=1$, 
then $\mbox{Btr}$ is simply the sum of all $n^2$~matrix entries \cite{devos}.  

\section{Example}

As an example, we choose the following U(6)~matrix:
\[
U = \frac{1}{12}\ \left( \begin{array}{cccccc}     
                     - 5 &    6 + 2i & - 5 - 5i & - 4 + 2i &   2      & - 2 -  i  \\
                  2 + 2i &  - 2 - 4i &     - 4i & - 3 -  i &   5 + 5i &   2 + 6i  \\
                - 6 - 3i &  - 2 - 2i &   1 + 3i & - 6      & - 4 - 2i &   3 + 4i  \\
                - 2 - 4i &  - 1 - 7i &   2 - 6i &   4 + 3i & - 1 - 2i &     - 2i  \\
                   3 - i &    4      & - 4 - 2i  &  2 - 2i & - 6 + 2i &   7 +  i  \\
                    - 6i &  - 1 + 3i & - 2 + 2i  &  3 + 6i &       5i & - 2 + 4i 
                         \end{array} \right) \ .
\]
Hence $n=6$. For $m$, we invesitigate all different possibilities, 
i.e.\ $m=1$, $m=2$, and $m=3$.
During the numerical procedure, the progress parameter~$\Psi$
diminishes according to Table~\ref{tabel1}.
We see that, after 36~iterations, $\Psi$~already approaches~0.
Therefore, below we give results for $\tau = 36$.

\begin{table}
\caption{Progress parameter $\Psi$ as a function of
         the number~$t$ of iteration steps.}
\begin{center}
\begin{tabular}{|r|rrr|}
\hline
    & & & \\[-2mm]
    & $m=1$ & $m=2$ & $m=3$ \\[2mm]
\hline
    & & & \\[-2.5mm]
$t$ & \multicolumn{3}{c|}{$\Psi_t$} \\[1.5mm]
\hline
    &        &        &        \\[-2mm]
 0  & 34.889 & 32.000 & 33.743 \\
 1  &  4.407 &  9.517 &  6.643 \\
 2  &  2.573 &  4.332 &  2.533 \\
 3  &  1.381 &  2.680 &  1.023 \\
 4  &  0.586 &  1.627 &  0.513 \\
 5  &  0.213 &  0.868 &  0.375 \\
 6  &  0.084 &  0.577 &  0.318 \\
 7  &  0.042 &  0.492 &  0.277 \\
 8  &  0.027 &  0.461 &  0.240 \\
 9  &  0.020 &  0.442 &  0.206 \\
10  &  0.016 &  0.423 &  0.174 \\
11  &  0.014 &  0.400 &  0.147 \\  
12  &  0.012 &  0.372 &  0.122 \\ 
13  &  0.010 &  0.339 &  0.101 \\ 
14  &  0.009 &  0.303 &  0.083 \\
15  &  0.008 &  0.264 &  0.067 \\
... &        &        &        \\
36  &  0.001 &  0.001 & 0.001  \\
    &        &        &        \\[1.5mm]
\hline
\end{tabular}
\end{center}
\label{tabel1}
\end{table}

We thus find, after 
36~iterations\footnote{Each iteration, in turn, 
                   needs $2r$~polar decompositions.
                   These are performed by Hero's iterative method
                   (a.k.a. Heron's method).
                   For each, we applied only ten iterations.}:
\begin{itemize}
\item for $m=1$:
      \[ \hspace*{-35mm}
      X = \left( \begin{array}{rrrrrr}
 0.27 - 0.31i & -0.27 + 0.45i &  0.58 + 0.13i &  0.28 - 0.24i &  0.07 + 0.15i &  0.07 - 0.17i \\  
 0.04 + 0.23i & -0.12 - 0.35i &  0.26 - 0.21i & -0.01 - 0.26i &  0.57 + 0.13i &  0.25 + 0.46i \\  
 0.51 + 0.22i &  0.23 + 0.06i & -0.02 - 0.26i &  0.42 + 0.27i &  0.27 - 0.25i & -0.42 - 0.03i \\  
 0.37 - 0.03i &  0.57 - 0.14i &  0.28 + 0.45i & -0.42 - 0.04i &  0.06 - 0.18i &  0.16 - 0.06i \\  
 0.26 + 0.01i &  0.33 - 0.04i & -0.16 - 0.33i &  0.23 + 0.05i & -0.23 + 0.47i &  0.57 - 0.16i \\  
-0.49 - 0.12i &  0.26 + 0.02i &  0.06 + 0.23i &  0.51 + 0.22i &  0.26 - 0.33i &  0.37 - 0.03i \end{array} \right) \ ,   
\]
      with the following row sums and column sums:
      \bea
      r_1 & = & 1.002 + 0.000i \\
      r_2 & = & 0.998 - 0.001i \\
      r_3 & = & 1.007 + 0.001i \\
      r_4 & = & 1.007 + 0.002i \\
      r_5 & = & 0.998 - 0.000i \\
      r_6 & = & 0.988 - 0.002i \\
      c_1 & = & 0.989 - 0.002i \\
      c_2 & = & 0.998 - 0.001i \\
      c_3 & = & 0.997 - 0.000i \\
      c_4 & = & 1.006 + 0.001i \\
      c_5 & = & 1.002 + 0.001i \\
      c_6 & = & 1.008 + 0.001i \ ;
      \eea 
\item for $m=2$:
      \[ \hspace*{-35mm}
      X = \left( \begin{array}{rrrrrr}
     0.33 - 0.05i &
    -0.39 - 0.07i &
     0.61 + 0.34i &
     0.32 + 0.19i &
     0.06 - 0.29i &
     0.08 - 0.12i
\\   
    -0.30 - 0.16i &
     0.35 + 0.36i &
     0.07 + 0.16i &
     0.08 + 0.09i &
     0.23 - 0.01i &
     0.57 - 0.45i
\\
     0.41 - 0.38i &
     0.51 - 0.02i &
     0.34 + 0.14i &
    -0.30 - 0.12i &
     0.26 + 0.24i &
    -0.21 + 0.13i
\\
     0.36 - 0.08i &
     0.26 - 0.27i &
    -0.17 - 0.26i &
     0.56 + 0.34i &
    -0.19 + 0.34i &
     0.17 - 0.07i
\\
     0.26 + 0.43i &
    -0.12 + 0.08i &
     0.06 - 0.48i &
    -0.02 - 0.06i &
     0.68 + 0.04i &
     0.13 - 0.02i
\\
    -0.06 + 0.24i &
     0.39 - 0.09i &
     0.10 + 0.09i &
     0.35 - 0.43i &
    -0.04 - 0.34i &
     0.26 + 0.52i
                 \end{array} \right) \ ,
      \]
      with row sums and column sums
      \bea
      r_1 & = & \left( \begin{array}{rr}  0.994 - 0.001i &  0.003 - 0.002i \\ 
                                          0.001 + 0.000i &  0.997 - 0.000i \end{array} \right) \\
      r_2 & = & \left( \begin{array}{rr}  1.003 + 0.002i & -0.001 - 0.005i \\ 
                                         -0.003 - 0.000i &  1.003 + 0.002i \end{array} \right) \\   
      r_3 & = & \left( \begin{array}{rr}  1.002 + 0.002i & -0.001 + 0.001i \\ 
                                          0.001 - 0.005i &  1.000 + 0.001i \end{array} \right) \\
      c_1 & = & \left( \begin{array}{rr}  1.002 + 0.001i & -0.004 - 0.002i \\ 
                                         -0.004 - 0.001i &  1.001 + 0.001i \end{array} \right) \\
      c_2 & = & \left( \begin{array}{rr}  1.002 + 0.001i &  0.001 + 0.002i \\ 
                                          0.002 - 0.001i &  0.998 - 0.001i \end{array} \right) \\
      c_3 & = & \left( \begin{array}{rr}  0.995 - 0.002i &  0.002 - 0.000i \\ 
                                          0.002 + 0.002i &  1.001 + 0.000i \end{array} \right) \ ;
      \eea
\item for $m=3$:
      \[ \hspace*{-30mm}
      X = \left( \begin{array}{rrrrrr}    
     0.54 + 0.37i &
    -0.10 - 0.25i &
     0.16 - 0.13i &
     0.45 - 0.36i &
     0.10 + 0.24i &
    -0.16 + 0.13i
\\
    -0.23 - 0.13i &
     0.47 - 0.06i &
    -0.07 - 0.41i &
     0.23 + 0.13i &
     0.53 + 0.06i &
     0.07 + 0.41i
\\
    -0.13 - 0.16i &
    -0.37 - 0.19i &
     0.53 + 0.17i &
     0.13 + 0.16i &
     0.37 + 0.19i &
     0.47 - 0.17i
\\
     0.46 - 0.37i &
     0.10 + 0.24i &    
    -0.16 + 0.13i &
     0.54 + 0.36i &
    -0.10 - 0.24i &
     0.16 - 0.13i
\\
     0.23 + 0.13i &
     0.53 + 0.06i &
     0.07 + 0.41i &
    -0.23 - 0.14i &
     0.47 - 0.06i &
    -0.07 - 0.41i
\\
     0.13 + 0.16i &
     0.37 + 0.19i &
     0.47 - 0.17i &
    -0.13 - 0.16i &
    -0.37 - 0.19i &
     0.52 + 0.17i
                 \end{array} \right) \ ,
      \]
      with row sums and column sums
      \bea
      r_1 & = & \left( \begin{array}{rrr}  0.997 + 0.001i & -0.004 - 0.001i & -0.002 - 0.002i \\ 
                                          -0.001 + 0.002i &  0.999 + 0.000i &  0.001 - 0.001i \\ 
                                           0.001 + 0.001i &  0.002 - 0.001i &  1.003 - 0.001i \end{array} \right) \\
      r_2 & = & \left( \begin{array}{rrr}  1.002 - 0.001i &  0.002 + 0.000i & -0.000 + 0.001i \\ 
                                           0.003 - 0.003i &  1.000 - 0.000i & -0.003 + 0.000i \\ 
                                           0.001 - 0.002i & -0.001 + 0.001i &  0.997 + 0.001i\end{array} \right) \\
      c_1 & = & \left( \begin{array}{rrr}  1.000 + 0.000i & -0.000 - 0.003i &  0.001 - 0.003i \\ 
                                           0.002 + 0.002i &  1.000 + 0.000i &  0.000 - 0.002i \\ 
                                           0.003 + 0.001i &  0.001 + 0.001i &  1.000 - 0.000i\end{array} \right) \\
      c_2 & = & \left( \begin{array}{rrr}  1.000 - 0.000i &  0.000 + 0.003i & -0.001 + 0.003i \\ 
                                          -0.001 - 0.002i &  1.000 - 0.000i & -0.000 + 0.002i \\ 
                                          -0.003 - 0.001i & -0.002 - 0.001i &  1.000 + 0.000i\end{array} \right) \ .
      \eea
\end{itemize}
For $m=2$, we also give the corresponding biunitary vector:
\[
U \left( \begin{array}{rr}  1.00 - 0.00i &  0.00 + 0.00i \\
                            0.00 - 0.00i &  1.00 + 0.00i \\[1mm]
                            0.81 - 0.31i &  0.23 + 0.43i \\
                           -0.20 + 0.44i &  0.84 + 0.25i \\[1mm]
                           -0.34 + 0.77i & -0.37 + 0.38i \\
                           -0.29 - 0.45i &  0.19 + 0.83i \end{array} \right) 
=
  \left( \begin{array}{rr} -0.95 - 0.16i &  0.24 - 0.14i \\
                           -0.14 - 0.24i & -0.91 - 0.31i \\[1mm]
                            0.06 - 0.70i & -0.71 + 0.01i \\
                           -0.28 - 0.65i &  0.62 - 0.34i \\[1mm]
                           -0.12 - 0.73i &  0.67 - 0.03i \\
                           -0.48 - 0.46i & -0.57 + 0.47i \end{array} \right) \ .
\]

\section{Permutation matrices}

Although we lack a proof of Conjecture~1 
in the case of an arbitrary unitary matrix~$U$,
we can say that Conjecture~1 is certainly true for the case where $U$
is an arbitrary $n \times n$ permutation matrix.
Indeed, any permutation matrix of size of $n \times n = mr \times mr$ 
can be decomposed as a product of 
three permutation matrices $D$, $X$, and $Z$,
the matrix~$D$ belonging to the group DU($n, m$),
the matrix~$X$ belonging to the group XU($n, m$), and 
the matrix~$Z$ belonging to the group ZU($n, m$).
In fact, 
$D$~belongs to a finite subgroup of DU($n, m$), of order $(m!)^r$     and
isomorphic to the product {\bf S}$_m^r$ of symmetric goups,
$X$~belongs to a finite subgroup of XU($n, m$), of order $(r!)^m$     and
isomorphic to the product {\bf S}$_r^m$ of symmetric goups, and
$Z$~belongs to a finite subgroup of ZU($n, m$), of order $(m!)^{r-1}$ and
isomorphic to the product {\bf S}$_m^{r-1}$.
The fact that such a decomposition is always possible 
\cite{devosvanrentergem1} \cite{devosvanrentergem2},
is a consequence of Birkhoff's theorem \cite{birkhoff} on doubly stochastic matrices
(with rational entries). 
The decomposition has been applied both 
in Clos networks of telephone switching systems \cite{clos} \cite{hwang} and
in reversible computing \cite{revc}. 

As an example, we choose the following $6 \times 6$ permutation matrix:
\[
U = \left( \begin{array}{rrrrrr} 0 & 0 & 0 & 0 & 1 & 0 \\
                                 1 & 0 & 0 & 0 & 0 & 0 \\
                                 0 & 1 & 0 & 0 & 0 & 0 \\
                                 0 & 0 & 0 & 1 & 0 & 0 \\
                                 0 & 0 & 0 & 0 & 0 & 1 \\
                                 0 & 0 & 1 & 0 & 0 & 0 \end{array} \right) \ . 
\]
For $m$, we investigate all different 
non-trivial\footnote{We note that, for permuation matrices, 
                     not only the case $m=n$ is trivial,
                     but also the case $m=1$: suffice it 
                     to choose both $D$ and $Z$ equal to the $n \times n$ unit matrix and 
                     to choose              $X$ equal to $U$.} 
possibilities: $m=2$ and $m=3$.
We have:
\begin{itemize}
\item for $m=2$:
      \[
      U = \left( \begin{array}{rrrrrr} 0 & 1 &   &   &   &   \\
                                       1 & 0 &   &   &   &   \\
                                         &   & 0 & 1 &   &   \\
                                         &   & 1 & 0 &   &   \\
                                         &   &   &   & 1 & 0 \\
                                         &   &   &   & 0 & 1 \end{array} \right)
          \left( \begin{array}{rrrrrr} 1 & 0 & \ 0 & 0 & \ 0 & 0 \\
                                       0 & 0 &   0 & 0 &   0 & 1 \\[1mm]
                                       0 & 0 &   1 & 0 &   0 & 0 \\
                                       0 & 1 &   0 & 0 &   0 & 0 \\[1mm]
                                       0 & 0 &   0 & 0 &   1 & 0 \\
                                       0 & 0 &   0 & 1 &   0 & 0 \end{array} \right)
          \left( \begin{array}{rrrrrr} 1 & 0 &   &   &   &   \\
                                       0 & 1 &   &   &   &   \\
                                         &   & 0 & 1 &   &   \\
                                         &   & 1 & 0 &   &   \\
                                         &   &   &   & 0 & 1 \\
                                         &   &   &   & 1 & 0 \end{array} \right) \ ,     
      \]
      where indeed the middle matrix has six unit line sums 
      $r_1=r_2=r_3=c_1=c_2=c_3={\tiny \left( \begin{array}{rr} 1 & \\ & 1 \end{array} \right)}$;
\item for $m=3$:
      \[
      U = \left( \begin{array}{rrrrrr} 0 & 0 & 1 &   &   &   \\
                                       1 & 0 & 0 &   &   &   \\
                                       0 & 1 & 0 &   &   &   \\
                                         &   &   & 1 & 0 & 0 \\
                                         &   &   & 0 & 1 & 0 \\
                                         &   &   & 0 & 0 & 1 \end{array} \right)
          \left( \begin{array}{rrrrrr} 1 & 0 & 0 & \ 0 & 0 & 0 \\
                                       0 & 1 & 0 &   0 & 0 & 0 \\
                                       0 & 0 & 0 &   0 & 0 & 1 \\[1mm]
                                       0 & 0 & 0 &   1 & 0 & 0 \\
                                       0 & 0 & 0 &   0 & 1 & 0 \\
                                       0 & 0 & 1 &   0 & 0 & 0 \end{array} \right)
          \left( \begin{array}{rrrrrr} 1 & 0 & 0 &   &   &   \\
                                       0 & 1 & 0 &   &   &   \\
                                       0 & 0 & 1 &   &   &   \\
                                         &   &   & 1 & 0 & 0 \\
                                         &   &   & 0 & 0 & 1 \\
                                         &   &   & 0 & 1 & 0 \end{array} \right) \ ,     
     \]      
     where indeed the middle matrix has four unit line sums 
     $r_1=r_2=c_1=c_2={\tiny \left( \begin{array}{rrr} 1 & & \\ & 1 & \\ & & 1 \end{array} \right)}$.
\end{itemize}

Because Conjecture~1 
is true for any $n \times n$         permutation matrix, it also 
is true for any $n \times n$ complex permutation matrix
(i.e.\ unitary matrix with only one non-zero entry in every row and column).
Such matrices form an $n$-dimensional non-connected subgroup of 
the $n^2$-dimensional group U($n$)
(consisting of $n!$ components, each $n$-dimensional).
We can indeed decompose such matrix as $D'P$, 
where $D'$ is a diagonal unitary matrix and $P$ is a permutation matrix.
We decompose $P$ as $D''XZ$, leading to the decomposition $D'D''XZ$
of the complex  permutation matrix.
Introducing $D=D'D''$, we obtain a desired decomposition $DXZ$.

\section{Conclusion}

Every $n \times n$ unitary matrix has an   Idel--Wolf decomposition.
If $n$ is even, then it also has a F\"uhr--Rzeszotnik decomposition.
We conjecture that, if $n$ is a composed integer,
it has as many similar decompositions as $n$ has divisors.
We offer no proof, as generalization of either
the Idel--Wolf         proof (based on symplectic topology) or 
the F\"uhr--Rzeszotnik proof (based on linear algebra) is not straightforward.
We provide an iterative algorithm for finding
a numerical approximation of each of the conjectured decompositions.
Finally, we demonstrate that the conjecture is true 
for $n \times n$ (complex) permutation matrices.


\section*{Appendix}

In Section 8, multiplying $X_{t-1}$ to the left with $L_t$ increases its block trace:
\[
     |\mbox{Btr}(L_tX_{t-1})| =
\left|\ \sum_{j=1}^r \sum_{k=1}^r \, \mbox{Tr}((L_tX_{t-1})_{jk})\ \right| =
\left|\ \sum_{j=1}^r \sum_{k=1}^r \, \mbox{Tr}(\Fie_j^{-1}\, (X_{t-1})_{jk})\ \right| 
\]
\[
= 
\left|\ \sum_{j=1}^r              \, \mbox{Tr}(P_j)\ \right| \ge
\left|\ \sum_{j=1}^r              \, \mbox{Tr}(\Fie_jP_j)\ \right| =
\left|\ \sum_{j=1}^r \sum_{k=1}^r \, \mbox{Tr}((X_{t-1})_{jk})\ \right| = 
|\mbox{Btr}(X_{t-1})| \ .
\]
Analogously, 
multiplying $L_tX_{t-1}$ to the right with $R_t$ increases its block trace.
Hence, we have
\[
|\mbox{Btr}(X_t)| = |\mbox{Btr}(L_tX_{t-1}R_t)| \ge |\mbox{Btr}(L_tX_{t-1})| 
                                                \ge |\mbox{Btr}(X_{t-1})|\ .
\]

The increasing value of $|\mbox{Btr}(X_t)|$ is bounded by the value~$n$.
An $n \times n$ unitary matrix $A$ has $|\mbox{Btr}(A)|=n$ 
iff it is a member of the group $e^{i\alfa}$~XU($n,m$). 
These two facts 
are proved by reasoning as in Appendix~A of De Vos and De Baerdemacker \cite{devos},
by considering the following property of the row sums~$r_a$ and column sums~$c_b$ of~$A$:
\[
\sum_{a=1}^r \sum_{j=1}^m  \sum_{k=1}^m \ \left|(r_a)_{jk}\right|^2 = 
\sum_{b=1}^r \sum_{j=1}^m  \sum_{k=1}^m \ \left|(c_b)_{jk}\right|^2 = n \ ,
\] 
a fact which, in turn, 
is  proved by reasoning as in Appendix~A of De Vos, Van Laer, and Vandenbrande \cite{vanlaer}.

\noindent
{\bf Acknowledgements}. 
SDB acknowleddes 
the Canada Research Chair program and
the New Brunswick Innovation Foundation.


\end{document}